\documentclass[11pt]{article}
\usepackage{graphicx}
\usepackage{amsmath}
\usepackage{amssymb}
\usepackage{caption2}
\begin{document}
\bibliographystyle{prsty}
\begin{center}
{\large {\bf \sc{  Analysis of the $X(1835)$ and related baryonium states with Bethe-Salpeter equation }}} \\[2mm]
Zhi-Gang Wang \footnote{E-mail,wangzgyiti@yahoo.com.cn.  }    \\
 Department of Physics, North China Electric Power University, Baoding 071003, P. R. China
\end{center}

\begin{abstract}
In this article, we study the mass spectrum of the baryon-antibaryon
bound states $p\bar{p}$, $\Sigma\bar{\Sigma}$, $\Xi\bar{\Xi}$,
$\Lambda\bar{\Lambda}$, $p\bar{N}(1440)$,
$\Sigma\bar{\Sigma}(1660)$, $\Xi\bar{\Xi}^\prime$ and
$\Lambda\bar{\Lambda}(1600)$  with the Bethe-Salpeter equation. The
numerical results indicate that the $p\bar{p}$,
$\Sigma\bar{\Sigma}$, $\Xi\bar{\Xi}$, $p\bar{N}(1440)$,
$\Sigma\bar{\Sigma}(1660)$, $\Xi\bar{\Xi}^\prime$ bound states maybe
exist, and the new resonances $X(1835)$ and $X(2370)$ can be
tentatively identified as the $p\bar{p}$ and $p\bar{N}(1440)$ (or
$N(1400)\bar{p}$) bound states respectively with some gluon
constituents, and the new resonance  $X(2120)$ may be a pseudoscalar
glueball. On the other hand, the Regge trajectory favors identifying
the  $X(1835)$, $X(2120)$ and $X(2370)$ as the excited
$\eta^\prime(958)$ mesons with the radial quantum numbers  $n=3$,
$4$ and $5$, respectively.

\end{abstract}



\section{Introduction}
In 2003, the BES collaboration  observed a significant narrow
near-threshold enhancement in the proton-antiproton ($p\bar{p}$)
invariant mass spectrum  in the radiative decay $J/\psi\to\gamma
p\overline{p}$ \cite{BES03}.
 The enhancement can be fitted  with either an $S$-wave or a
 $P$-wave Breit-Wigner
resonance function. In the case of the $S$-wave fitted form, the
 mass and the width are $M = \left(1859 {}^{+3}_{-10}
{}^{+5}_{-25} \right)\,\rm{MeV}$ and $\Gamma < 30 \,\rm{MeV}$,
respectively. In 2005, the BES collaboration observed a resonance
state $X(1835)$ in the $\eta^{\prime}\pi^+\pi^-$ invariant mass
spectrum in the process $J/\psi\to\gamma\pi^{+}\pi^{-}\eta^{\prime}$
with the Breit-Wigner mass $M=(1833.7\pm 6.2\pm 2.7)\,\rm{MeV}$  and
the width $\Gamma=(67.7\pm 20.3\pm 7.7)\,\rm{MeV}$, respectively
\cite{BES05}. Recently,  the $X(1835)$ was confirmed by the BES
collaboration in the radiative decay $J/\psi\to\gamma \pi^+\pi^-
\eta'$ with  a statistical significance larger than $20\,\sigma$,
the fitted mass and width are
$M=\left(1836.5\pm3.0^{+5.6}_{-2.1}\right)\,\rm{MeV}$  and $\Gamma=
\left(190 \pm 9^{+38}_{-36}\right)\,\rm{MeV}$,  respectively
\cite{BES-Shen,BES1012}. The mass is consistent with the BESII
result \cite{BES05}, while the width is significantly larger.
Furthermore, the BES collaboration observed two new resonances
$X(2120)$ and $X(2370)$  in the $ \pi^+\pi^- \eta'$ invariant mass
spectrum with statistical significances larger than $7.2\,\sigma$
and $6.4\,\sigma$, respectively. The measured masses and widths are
$M_{X(2120)}=\left(2122.4\pm 6.7^{+4.7}_{-2.7}\right)\,\rm{MeV}$,
$M_{X(2370)}=\left(2376.3\pm8.7^{+3.2}_{-4.3}\right)\,\rm{MeV}$,
$\Gamma_{X(2120)}= (83 \pm 16^{+31}_{-11})\,\rm{MeV}$ and
$\Gamma_{X(2370)}=(83\pm17^{+44}_{-6})\,\rm{MeV}$, respectively
\cite{BES-Shen,BES1012}.

Many theoretical works were stimulated to interpret the nature and
the structure of the new resonance  $X(1835)$, such as the
$p\bar{p}$ bound state \cite{pp-X}, the pseudoscalar glueball
\cite{gb-Kochelev-PLB,gb-Hao}, and the radial excitation of the
$\eta'$ \cite{eta-Huang,eta-Klempt,eta-Li}, the threshold cusp
\cite{cusp-bugg}, etc.

In Ref.\cite{Ding2010}, Liu, Ding and Yan  study the decay widths of
 the second and the third radial excitations of the pseudoscalar mesons
  $\eta$ and $\eta'$
with the $^3P_0$ model, and observe  that the interpretation of the
$\eta(1760)$ and $X(1835)$ as the second radial excitations  of the
$\eta$ and $\eta'$ crucially depends on the measured mass and width
of the $\eta(1760)$, and suggest that there may be sizable
$p\bar{p}$ content in the $X(1835)$, and the $X(2120)$ and $X(2370)$
cannot be understood as the third radial excitations of the $\eta$
and $\eta'$ respectively, the $X(2370)$ probably is a mixture of the
$\eta'(4{^1}{S}{_0})$ and glueball.

In Ref.\cite{WangCPL},  we take  the $X(1835)$  as a baryonium state
with the quantum numbers $J^{PC}=0^{-+}$, and calculate the mass
spectrum of the baryon-antibaryon bound states $p\bar{p}$,
$\Sigma\bar{\Sigma}$, $\Xi\bar{\Xi}$, and $\Lambda\bar{\Lambda}$ in
the framework of the Bethe-Salpeter equation with a phenomenological
potential. The numerical results indicate that the $p\bar{p}$,
$\Sigma\bar{\Sigma}$ and $\Xi\bar{\Xi}$ bound states maybe exist,
and the $X(1835)$ can be tentatively identified as the $p\bar{p}$
bound state. In this article, we extend our previous work to study
whether or not there exist the baryon-antibaryon bound states
$p\bar{N}(1440)$, $\Sigma\bar{\Sigma}(1660)$, $\Xi\bar{\Xi}^\prime$
and $\Lambda\bar{\Lambda}(1600)$ (or $N(1440)\bar{p}$,
$\Sigma(1660)\bar{\Sigma}$, $\Xi^\prime\bar{\Xi}$ and
$\Lambda(1600)\bar{\Lambda}$), here the $\prime$ denotes the first
radial excited state.  In the scenario of the coupled channel
effects or the hadronic dressing mechanism \cite{HDress}, the
pseudoscalar mesons $X(1835)$, $X(2120)$ and $X(2370)$ can be  taken
as having small pseudoscalar $q\bar{q}$ kernels of the typical
$q\bar{q}$ meson size. The strong couplings to the virtual
intermediate hadronic states (for example, the $p\bar{p}$,
$p\bar{N}(1440)$, $\Sigma\bar{\Sigma}$, $\Xi\bar{\Xi}$, etc) can
change the bare masses which originate from the quark-gluon
interactions significantly, and enrich the pure $q\bar{q}$ states
with some baryon-antibaryon components. In the present case, the
input parameters of the Bethe-Salpeter equation should be readjusted
to obtain reasonable predicted masses if the coupled-channel effects
are considered. However, it is beyond the scope of the present work.
 Such a scenario needs
detailed studies.

The article is arranged as follows:  we solve the Bethe-Salpeter
equation  for the baryon-antibaryon bound  states  in Sec.2; in
Sec.3, we present the numerical results and discussions; and Sec.4
is reserved for our conclusions.

\section{Bethe-Salpeter equation}
The Bethe-Salpeter equation  is a conventional approach in dealing
with the two-body relativistic bound state problems,  and has given
many successful descriptions of the hadron  properties
\cite{BS51,BS-Roberts}. We write down the ladder Bethe-Salpeter
equation for  the pseudoscalar  bound states in the Euclidean
spacetime\footnote{In this article, we use the metric
$\delta_{\mu\nu}=(1,1,1,1)$, $\left\{\gamma_\mu
\gamma_\nu+\gamma_\nu \gamma_\mu\right\}=2\delta_{\mu\nu}$, the
momentums $k_\mu=(k_4,\overrightarrow{k})$,
$q_\mu=(q_4,\overrightarrow{q})$ and $P_\mu=(iE,\overrightarrow{P})$
with $P^2=-M^2_{X}$.}, which can be derived from the Euclidean
path-integral formulation of the theory,
\begin{eqnarray}
  S_1^{-1}\left(q+\xi_1P\right)\chi(q,P)S_2^{-1}\left(q-\xi_2P\right)&=&\int\frac{d^4 k}{(2\pi)^4}\gamma_5 \chi(k,P) \gamma_5 G(q-k)\, ,\\
  S_{1/2}^{-1}\left(q\pm \xi_{1/2}P\right)&=&i\left(\gamma \cdot q\pm \xi_{1/2} \gamma \cdot P\right)+M_{1/2} \, , \nonumber\\
  \xi_{1/2}&=&\frac{M_{1/2}}{M_1+M_2} \, , \nonumber
\end{eqnarray}
where the $P_\mu$ is the four-momentum of the center of mass of the
 baryon-antibaryon bound state, the $q_\mu$ is the relative
four-momentum between the baryon and antibaryon,
 $\gamma_{5}$ is the bare baryon-meson vertex,
the $\chi(q,P)$ is the Bethe-Salpeter amplitude of the
 baryon-antibaryon bound state, and the $G(q-k)$ is the interaction
kernel.

In the flavor $SU(3)$ symmetry limit,  the interactions among the
ground state octet baryons and the pseudoscalar mesons can be
described by the lagrangian $\cal L$,
\begin{eqnarray}
{\cal L} = \sqrt2 \left( D {\rm Tr} \left(\bar B
\left\{P,B\right\}_{+}\right)+ F {\rm Tr} \left(\bar B \left[ P,B
\right]_{-} \right) \right)\, ,
\end{eqnarray}
where
\begin{eqnarray}
B &=& \left(
\begin{array}{ccc}
    \frac{1}{\sqrt{2}} \Sigma^0 + \frac{1}{\sqrt{6}} \Lambda & \Sigma^+ & p \\
    \Sigma^- & - \frac{1}{\sqrt{2}}\Sigma^0 + \frac{1}{\sqrt{6}} \Lambda & n \\
    \Xi^- & \Xi^0 & - \frac{2}{\sqrt{6}} \Lambda
\end{array} \right) \, ,\nonumber\\
P &=& \left(
\begin{array}{ccc}
    \frac{1}{\sqrt2} \pi^0 + \frac{1}{\sqrt{6}}\eta & \pi^+ & K^+ \\
    \pi^- & - \frac{1}{\sqrt2} \pi^0 + \frac{1}{\sqrt6} \eta & K^0 \\
    K^- & \bar K^0  &- \frac{2}{\sqrt6} \eta
\end{array}
\right) \, ,
\end{eqnarray}
and the $D$ and $F$ are two parameters for the coupling constants.
From the lagrangian, we can obtain
\begin{eqnarray}
g_{\pi^0 p p} &=& -g_{\pi^0 n n}=D + F\, , \,\, g_{\pi^0 \Sigma^+
\Sigma^+} =- g_{\pi^0 \Sigma^- \Sigma^-} =2 F\, ,
\nonumber \\
g_{\pi^0 \Xi^- \Xi^-} &=& -g_{\pi^0 \Xi^0 \Xi^0}=D-F\, , \,\,
g_{\eta p p} = g_{\eta nn}=-\frac{D-3F}{\sqrt3} \, ,
\nonumber \\
g_{\eta \Sigma^+ \Sigma^+} &=& g_{\eta \Sigma^- \Sigma^-} =g_{\eta
\Sigma^0 \Sigma^0} =- g_{\eta \Lambda \Lambda}=\frac{2D}{\sqrt3}
\, , \nonumber \\
g_{\eta \Xi^-\Xi^-} &=& g_{\eta \Xi^0\Xi^0}=-\frac{D+3F}{\sqrt3}\, ,
\end{eqnarray}
and write down the kernel $G(q-k)$ explicitly,
\begin{eqnarray}
G(q-k)&=&\frac{g^2(q-k)C_\pi}{(q-k)^2+m_\pi^2}+\frac{g^2(q-k)C_\eta}{(q-k)^2+m_\eta^2}
\, ,
\end{eqnarray}
where the coefficients  $C_\pi=(1+\alpha)^2$, $4\alpha^2$,
$(1-\alpha)^2$, $0$ and $C_\eta=\frac{(1-3\alpha)^2}{3}$,
$\frac{4}{3}$, $\frac{(1+3\alpha)^2}{3}$, $\frac{4}{3}$ for the
$p\bar{p}$, $\Sigma\bar{\Sigma}$, $\Xi\bar{\Xi}$,
$\Lambda\bar{\Lambda}$ bound states respectively; $g^2(k)=D^2$ and
$\alpha=\frac{F}{D}$.

With a simple replacement
\begin{eqnarray}
B\to B^\prime \, , \, D \to D^\prime \, , \, F \to F^\prime\, , \,
\alpha \to \alpha^\prime \, , \, g \to g^\prime \, ,
\end{eqnarray}
where
\begin{eqnarray}
B^\prime &=& \left(
\begin{array}{ccc}
    \frac{1}{\sqrt{2}} \Sigma(1660) + \frac{1}{\sqrt{6}} \Lambda(1600) & \Sigma(1660) & N(1440) \\
    \Sigma(1660) & - \frac{1}{\sqrt{2}}\Sigma(1660) + \frac{1}{\sqrt{6}} \Lambda(1600) & N(1440) \\
    \Xi^\prime(?) & \Xi^\prime(?) & - \frac{2}{\sqrt{6}} \Lambda(1600)
\end{array} \right)  \, ,
\end{eqnarray}
we can obtain the corresponding  couplings among the first radial
excited  octet baryons and the pseudoscalar mesons. The
corresponding coefficients are $C_\pi=(1+\alpha)(1+\alpha')$,
$4\alpha\alpha'$, $(1-\alpha)(1-\alpha')$, $0$ and
$C_\eta=\frac{(1-3\alpha)(1-3\alpha')}{3}$, $\frac{4}{3}$,
$\frac{(1+3\alpha)(1+3\alpha')}{3}$, $\frac{4}{3}$ for the bound
states $p\bar{N}(1440)$, $\Sigma\bar{\Sigma}(1660)$,
$\Xi\bar{\Xi}^\prime$, $\Lambda\bar{\Lambda}(1600)$ (or
$N(1440)\bar{p}$, $\Sigma(1660)\bar{\Sigma}$, $\Xi^\prime\bar{\Xi}$,
 $\Lambda(1600)\bar{\Lambda}$), respectively. In this article, we
can take the approximation $\alpha=\alpha'$ and $g=g'$ (i.e. we
assume that  the ground state and first radial excited octet baryons
have the same quantum numbers except for the masses), and use the
parameters determined in our previous work \cite{WangCPL}.
Furthermore, we also solve the Bethe-Salpeter equations with the
parameters $g'\neq g$, as the flavor $SU(3)$ symmetry does not
warrant $g'\approx g$.

In this article, we have neglected the baryon-antibaryon
annihilation effects. There are two typical diagrams for the
annihilation contributions, the exchanges of the intermediate gluons
and the intermediate mesons. The contributions of the intermediate
gluons $\propto \alpha_s^n$ with $n\geq 2$, where
$\alpha_s=\frac{g_s^2}{4\pi}$, the $g_s$ is the quark-gluon coupling
constant, the values of the $\alpha_s^n$ are usually very small and
can be neglected for the heavy-quark systems \cite{Godfrey1985}.
While in the light-quark systems, the gluonic annihilation effects
are expected to play a more important role. However, even in the
light mesons, the annihilation Hamiltonian is usually small, and
there is considerable phenomenological evidence to reinforce one's
expectation that it becomes weaker as a meson system becomes more
excited \cite{Godfrey1985}. The effects are somewhat alleviated by
this factor, thus in practice the gluonic annihilation effects are
often ignored. The annihilations to the intermediate mesons can be
estimated as
$\propto\left(\frac{g^2(M^2_X)}{M_X^2+m_{\pi,\eta,\cdots}^2}\right)^n$
with $n\geq1$. If we take the typical momentum
$q=\sqrt{2M_r|E|}\approx200\,\rm{MeV}$, where the $M_r$ is the
reduced mass in the $p\bar{p}$ system, and the $E$ is the bound
energy of the $X(1835)$ as the $p\bar{p}$ bound state, $M_X\gg q$,
and
$\left(\frac{g^2(M^2_X)}{M_X^2+m_{\pi,\eta,\cdots}^2}\right)^n\ll
\frac{g^2(q^2)}{q^2+m_{\pi,\eta,\cdots}^2}$, such  annihilation
effects can be neglected. The annihilation effects in other channels
can also be neglected with  analogous  arguments.

In Ref.\cite{WangCPL}, we choose the value $\alpha=0.6$ from the
analysis of the hyperon semi-leptonic decays \cite{FD}, and take the
coupling constant $g^2(k)$ as a modified Gaussian distribution
$g^2(k)=A \left(\frac{k^2}{\mu^2}\right)^2\exp\left(-
\frac{k^2}{\mu^2}\right)$ in the Euclidean spacetime,  where the
strength  $A$ and the distribution width $\mu$ are two free
parameters. The ultraviolet behavior of the modified Gaussian
distribution warrants the integral in the Bethe-Salpeter equation is
convergent.

 The Euclidean Bethe-Salpeter amplitude of the pseudoscalar baryon-antibaryon
  bound state can be decomposed as \cite{BS-Roberts}
 \begin{eqnarray}
 \chi(q,P)&=& \gamma_5\left\{F(q,P)+i\!\not\!{P}F_1(q,P) +i\!\not\!{q}F_2(q,P)+\left[\!\not\!{q},\!\not\!{P}\right]F_3(q,P)
             \right\} \, .
 \end{eqnarray}
As in our previous work, we take the approximation
\begin{eqnarray}
 \chi(q,P)&=&
 \gamma_5\left\{F(q,P)+i\!\not\!{P}F_1(q,P)\right\} \, ,
 \end{eqnarray}
 for simplicity.
We can denote the baryon fields  as $\Psi(x)$,  and perform the
  Fierz re-ordering to study the contributions from
different spinor  structures\footnote{Here we choose  the $X(1835)$
as an example to illustrate the estimation, and use the metric in
the Minkowski spacetime.},
 \begin{eqnarray}
\Psi_\alpha(0) \bar{\Psi}_\beta(x)&=&-\frac{1}{4}
\delta_{\alpha\beta}\bar{\Psi}(x)\Psi(0)
-\frac{1}{4}(\gamma^\mu)_{\alpha\beta}\bar{\Psi}(x)\gamma_\mu
\Psi(0) -\frac{1}{8}(\sigma^{\mu\nu})_{\alpha\beta}\bar{\Psi}(x)\sigma_{\mu\nu}\Psi(0) \nonumber\\
&&+\frac{1}{4}(\gamma^\mu
\gamma_5)_{\alpha\beta}\bar{\Psi}(x)\gamma_\mu \gamma_5
\Psi(0)+\frac{1}{4}(i \gamma_5)_{\alpha\beta}\bar{\Psi}(x)i \gamma_5
\Psi(0) \, .
\end{eqnarray}
The pseudoscalar current $\bar{\Psi}(x)i \gamma_5 \Psi(0)$, the
axialvector current $\bar{\Psi}(x)\gamma_\mu \gamma_5 \Psi(0)$ and
the tensor current $\bar{\Psi}(x)\sigma_{\mu\nu}\Psi(0)$ have
nonvanishing couplings with the pseudoscalar meson $X(1835)$,
\begin{eqnarray}
 i\gamma_5\langle0|\bar{\Psi}(x)i \gamma_5 \Psi(0)|X(P)\rangle &\propto& \gamma_5 A_0+ \gamma_5 q\cdot P A_1+\cdots\,,\nonumber\\
 \gamma^\mu\gamma_5\langle0|\bar{\Psi}(x)\gamma_\mu \gamma_5 \Psi(0)|X(P)\rangle  &\propto&\gamma_5\!\not\!{P} B_0+\gamma_5\!\not\!{q} B_1+\cdots\, , \\
 \sigma^{\mu\nu}\langle0|\bar{\Psi}(x)\sigma_{\mu\nu}\Psi(0)|X(P)\rangle&\propto&\gamma_5\left[\!\not\!{q},\!\not\!{P}\right]C_0+\cdots\,
 ,
\end{eqnarray}
the coefficients $A_0$, $A_1$, $B_0$, $B_1$, $C_0$ are functions of
the $q_\mu$ and $P_\mu$. We can take the estimations $P_\mu\sim
M_X\approx1.8\,\rm{GeV}$,  $q_\mu\sim
\sqrt{2M_r|E|}\approx0.2\,\rm{GeV}$,
$\left[\!\not\!{q},\!\not\!{P}\right]\sim
\sqrt{2M_r|E|}M_X\approx0.36\,\rm{GeV}^2$,   translate the $P_\mu$
and $q_\mu$ into dimensionless quantities, $P_\mu \to
\widetilde{P}_\mu \widetilde{\Lambda}$, $q_\mu \to \widetilde{q}_\mu
\widetilde{\Lambda}$ with $\widetilde{\Lambda}=1\,\rm{GeV}$, and
absorb the $\widetilde{\Lambda}$ into the coefficients $A_0$, $A_1$,
$B_0$, $B_1$, $C_0$, then $\widetilde{P}_\mu\sim 1.8$,
$\widetilde{q}_\mu\sim 0.2$,
$\left[\widetilde{\!\not\!{q}},\widetilde{\!\not\!{P}}\right]\sim
0.36$. Compared  with the term $\gamma_5 \!\not\!{P}$, the term
$\gamma_5 \!\not\!{q}$ is greatly suppressed and can be neglected.
Furthermore, we expect that couplings of the tensor currents to the
pseudoscalar mesons are weaker than that  of the pseudoscalar and
axialvector currents, and neglect the term
$\gamma_5\left[\!\not\!{q},\!\not\!{P}\right]$, which is also
suppressed as the $\gamma_5 \!\not\!{q}$.

 The Bethe-Salpeter amplitudes $F(q,P)$ and $F_{1}(q,P)$ can be expanded  in terms of
Tchebychev polynomials $T^{\frac{1}{2}}_{n}(\cos \theta)$
\cite{Guth},
\begin{eqnarray}
 F(q,P)&=&\sum_{n=0}^{\infty}i^nF^{n}(q^2,P^2) q^n P^n T^{\frac{1}{2}}_{n}(\cos \theta)\,\, , \nonumber \\
 F_{1}(q,P)&=&\sum_{n=0}^{\infty}i^nF_{1}^{n}(q^2,P^2) q^nP^nT^{\frac{1}{2}}_{n}(\cos \theta)\,\, ,
\end{eqnarray}
where  $\theta$ is  the included  angle between  $q_\mu$ and
$P_\mu$. If we translate the momenta $q_\mu$ and $P_\mu$ into the
dimensionless quantities $\widetilde{q}_\mu$ and $\widetilde{P}_\mu$
respectively, and absorb the $\widetilde{\Lambda}$ into the
$F^{n}(q^2,P^2)$ and $F_1^{n}(q^2,P^2)$, then
$\widetilde{q}\widetilde{P}T^{\frac{1}{2}}_{1}(\cos \theta)\sim
0.36\cos\theta\sim0$,
$\widetilde{q}^2\widetilde{P}^2T^{\frac{1}{2}}_{2}(\cos \theta)\sim
0.13\cos2\theta\sim-0.03$, here we have taken the average  $\cos
n\theta\approx\frac{1}{\pi}\int_0^{\pi}{\cos n\theta}{\sin^2\theta}
d\theta$.  It is impossible to solve an infinite series of coupled
equations of the $F^n(q^2,P^2)$ and $F_{1}^n(q^2,P^2)$, we have to
make truncation in one or the other ways. In this article, we
neglect the small terms with $n\geq 1$. Numerical calculations
indicate that taking only the terms with $n=0$  can give
satisfactory results. If we take into account the small terms with
$n\geq 1$, the predictions may be improved mildly.  In the
following, we will smear the index $0$ for simplicity.

 Multiplying both sides of the Bethe-Salpeter equation   by
 $\gamma_5\left[\!\not\!{q},\!\not\!{P}\right]$ and carrying out  the trace
 in the Dirac spinor space, we can obtain an  simple relation
 $F=(M_1+M_2)F_1$, the amplitudes $F(q^2,P^2)$ and $F_1(q^2,P^2)$ are not independent.
 The Bethe-Salpeter  amplitude can be written  as
\begin{eqnarray}
 \chi(q,P)&=&
 \gamma_5\left(1+\frac{i\!\not\!{P}}{M_1+M_2}\right)F(q^2,P^2)  \, ,
 \end{eqnarray}
 and the Bethe-Salpeter equation can be projected into the following form,
 \begin{eqnarray}
 \left( q^2+M_1M_2+\frac{M_1M_2}{(M_1+M_2)^2}P^2\right)F(q^2,P^2)&=&\int \frac{d^4k}{(2\pi)^4}
  F(k^2,P^2)G(q-k) \, .
 \end{eqnarray}

We can introduce a parameter $\lambda(P^2)$ and solve  above
equation as an eigenvalue problem.  If there really exists  a bound
state in the pseudoscalar channel, the mass of the bound state $X$
can be determined by the condition $\lambda(P^2=-M_{X}^2)=1$,
\begin{eqnarray}
\left(q^2+M_1M_2+\frac{M_1M_2}{(M_1+M_2)^2}P^2\right)F(q^2,P^2)&=&\lambda(P^2)\int
\frac{d^4k}{(2\pi)^4} F(k^2,P^2)G(q-k) \, .
\end{eqnarray}
If we take $q^2=0$ and assume  that there exists a physical
solution, then
\begin{eqnarray}
 \left(M_1M_2-\frac{M_1M_2}{(M_1+M_2)^2}M_X^2\right)F(0,-M_X^2)&=&\int \frac{d^4k}{(2\pi)^4} F(k^2,-M_X^2)G(0-k) \, .
\end{eqnarray}
In numerical calculations, we observe that the Bethe-Salpeter
amplitude $F(k^2,-M_X^2)$ has the same sign in the region
$k^2\geq0$,
\begin{eqnarray}
 M_1M_2-\frac{M_1M_2}{(M_1+M_2)^2}M_X^2&=&\int\frac{d^4k}{(2\pi)^4}\frac{F(k^2,-M_X^2)}{F(0,-M_X^2)}G(0-k) >0\, ,
\end{eqnarray}
 and obtain an simple relation (or constraint),
\begin{eqnarray}
M_X^2<(M_1+M_2)^2 \, ,
\end{eqnarray}
which survives for $q^2>0$ (although the relation is not explicit
for $q^2>0$), i.e. the bound energy $E_X$ originates from the
interacting kernel $G(k)$ and should be negative,
$E_X=M_X-M_1-M_2<0$. On the other hand, if the Bethe-Salpeter
amplitude $F(k^2,-M_X^2)$ changes sign in the region $k^2\geq0$,
which does not warrant the positive value
$\int\frac{d^4k}{(2\pi)^4}\frac{F(k^2,-M_X^2)}{F(0,-M_X^2)}G(0-k)
>0$, and the relation $M_X^2<(M_1+M_2)^2$ fails  to  survive.

\section{Numerical results and discussions}
The input parameters are taken as $m_\pi=135\,\rm{MeV}$,
$m_\eta=548\,\rm{MeV}$, $M_p=938.3\,\rm{MeV}$,
$M_{\Sigma^+}=1189.4\,\rm{MeV}$, $M_{\Xi^-}=1321.7\,\rm{MeV}$,
$M_{\Lambda}=1115.7\,\rm{MeV}$,
$M_{N(1440)}=(1420-1470)\,\rm{MeV}\approx1440\,\rm{MeV}$,
$M_{\Sigma(1660)}=(1630-1690)\,\rm{MeV}\approx1660\,\rm{MeV}$,
$M_{\Lambda(1600)}=(1560-1700)\,\rm{MeV}\approx1600\,\rm{MeV}$, and
$M_{X(1835)}=1833.7\,\rm{MeV}$ from the Review of Particle Physics
\cite{PDG}. The first radial excited state of the $\Xi$ has not been
observed yet, we take the approximation
$M_{\Xi^\prime}=M_\Xi+\frac{M_{\Sigma(1660)}-M_{\Sigma}+M_{N(1440)}-M_{p}+M_{\Lambda(1600)}-M_{\Lambda}}{3}\approx
M_{\Xi}+486\,\rm{MeV}$.  The strength $A$ and the distribution width
$\mu$ are  free parameters, we take the values $A=215$ and
$\mu=200\,\rm{MeV}$ for the $p\bar{p}$ bound state as in
Ref.\cite{WangCPL}, and  take the simple replacements
$\mu\rightarrow \mu \frac{M^2_{\Sigma}}{M_p^2}$, $ \mu
\frac{M^2_{\Xi}}{M_p^2}$ and $ \mu \frac{M^2_{\Lambda}}{M_p^2}$ to
take into account the flavor $SU(3)$ breaking effects for the
$\Sigma\bar{\Sigma}$ ($\Sigma\bar{\Sigma}(1660)$), $\Xi\bar{\Xi}$
 ($\Xi\bar{\Xi}')$ and $\Lambda\bar{\Lambda}$
($\Lambda\bar{\Lambda}(1600)$) bound states,  respectively.

We solve the Bethe-Salpeter equations  as  an eigen-problem
numerically by direct iterations, and observe that  the convergent
behaviors  are  very good. For the $p\bar{p}$, $\Sigma\bar{\Sigma}$,
$\Xi\bar{\Xi}$,  $p\bar{N}(1440)$, $\Sigma\bar{\Sigma}(1660)$ and
$\Xi\bar{\Xi}^\prime$  bound states, there
 exists  a solution with $\lambda(P^2=-M_{X}^2)=1$ and $E_X<0$.
 On the other hand, we cannot obtain a solution to satisfy the condition
  $\lambda(P^2=-M_{X}^2)=1$ for the $\Lambda\bar{\Lambda}$ and $\Lambda\bar{\Lambda}(1600)$ bound states.
  Experimentally, there are $\Lambda\bar{\Lambda}$ near threshold enhancements in the
decays $B^+\to \Lambda\bar{\Lambda}K^+$, $B^0\to
\Lambda\bar{\Lambda}K^0,\Lambda\bar{\Lambda}K^{*0}$
\cite{LambdaTH-1,LambdaTH-2} and above threshold enhancements in the
decays  $B^+\to \Lambda\bar{\Lambda}K^+$ \cite{LambdaTH-3} from the
Belle collaboration, the above threshold enhancements can be
identified as the $J/\psi$ and $\eta_c$ mesons respectively, the
decays  $J/\psi \to \Lambda\bar{\Lambda}$ and $\eta_c \to
\Lambda\bar{\Lambda}$ are observed. The $\Lambda\bar{\Lambda}$ near
threshold enhancement may be a $\Lambda\bar{\Lambda}$ baryonium
state or just a final-state re-scattering effect, more experimental
data are still needed to identify it. We can study the
baryon-antibaryon scattering amplitudes in unitary Chiral
perturbation theory by taking into account the intermediate
multichannel baryon-loops (for example, the $p\bar{p}$,
$\Lambda\bar{\Lambda}$, $\Sigma\bar{\Sigma}$, etc.), and adjust the
parameters in the phenomenological lagrangian to reproduce the
$p\bar{p}$ baryonium state $X(1835)$, and explore whether or not
there exists a pole related with the $\Lambda\bar{\Lambda}$
baryonium state.

\begin{table}
\begin{center}
\begin{tabular}{|c|c|c|c|c|c|c|}
\hline\hline
                            &$p\bar{p}$  & $\Sigma\bar{\Sigma}$ & $\Xi\bar{\Xi}$ & $p\bar{N}(1440)$  & $\Sigma\bar{\Sigma}(1660)$  & $\Xi\bar{\Xi}^\prime$  \\ \hline
  $M_X[\rm{MeV}]$           & $1833.7$   & $2317.8$             & $2612.4$       & $2344.0/2374.0^*$ & $2798.4/2828.4^*$           & $3103.4$   \\ \hline
  $E_X[\rm{MeV}]$           & $-42.9$    & $-61.0$              & $-31.0$        & $-34.3$           & $-51.0$                     & $-26.0$     \\ \hline
  ${\rm {expt}}[\rm{MeV}]$  & ?\,1836.5  & ?\,2376.3            & ?              &  ?\,2376.3        &?                            & ?           \\ \hline
 \hline
\end{tabular}
\end{center}
\caption{ The masses $M_X$ and bound energies $E_X$   of the
baryon-antibaryon  bound states, the $*$ denotes that the upper
bounds of the masses of the  $N(1440)$ and $\Sigma(1660)$ baryons
are taken. }
\end{table}

\begin{table}
\begin{center}
\begin{tabular}{|c|c|c|c|}
\hline\hline
                                                 &$p\bar{N}(1440)$       &$\Sigma\bar{\Sigma}(1660)$& $\Xi\bar{\Xi}^\prime$\\ \hline
 $M_X[\rm{MeV}]$ $(\mu=200\,\rm{MeV},\tau=0.55)$ &   $2378.2/2408.2^*$   &                          &         \\ \hline
 $M_X[\rm{MeV}]$ $(\mu=200\,\rm{MeV},\tau=0.61)$ &   $2375.3/2405.3^*$   & $2849.3/2879.3^*$        &         \\ \hline
 $M_X[\rm{MeV}]$ $(\mu=200\,\rm{MeV},\tau=0.78)$ &   $2363.2/2393.2^*$   & $2832.9/2862.9^*$        & $3129.3$ \\ \hline
 $M_X[\rm{MeV}]$ $(\mu=200\,\rm{MeV},\tau=0.90)$ &   $2353.2/2383.2^*$   & $2813.9/2843.9^*$        & $3118.1$ \\ \hline
 $M_X[\rm{MeV}]$ $(\mu=200\,\rm{MeV},\tau=1.00)$ &   $2344.0/2374.0^*$   & $2798.4/2828.4^*$        & $3103.4$ \\ \hline
 $M_X[\rm{MeV}]$ $(\mu=400\,\rm{MeV},\tau=0.44)$ &   $2378.2/2408.2^*$   &                          &         \\ \hline
 $M_X[\rm{MeV}]$ $(\mu=400\,\rm{MeV},\tau=0.47)$ &   $2373.2/2403.2^*$   & $2849.3/2879.3^*$        &         \\ \hline
 $M_X[\rm{MeV}]$ $(\mu=400\,\rm{MeV},\tau=0.52)$ &   $2361.8/2391.8^*$   & $2832.3/2862.3^*$        & $3129.3$ \\ \hline
 $M_X[\rm{MeV}]$ $(\mu=400\,\rm{MeV},\tau=0.60)$ &   $2336.8/2366.8^*$   & $2784.4/2814.4^*$        & $3084.4$ \\ \hline
 ${\rm {expt}}[\rm{MeV}]$                        &     ?\,2376.3         &?                         & ?        \\ \hline
  \hline
\end{tabular}
\end{center}
\caption{ The masses $M_X$ of the baryon-antibaryon  bound states
with variations of the input parameters, the $*$ denotes that the
upper bounds of the masses of the  $N(1440)$ and $\Sigma(1660)$
baryons are taken. }
\end{table}

\begin{figure}
 \centering
 \includegraphics[totalheight=7cm,width=9cm]{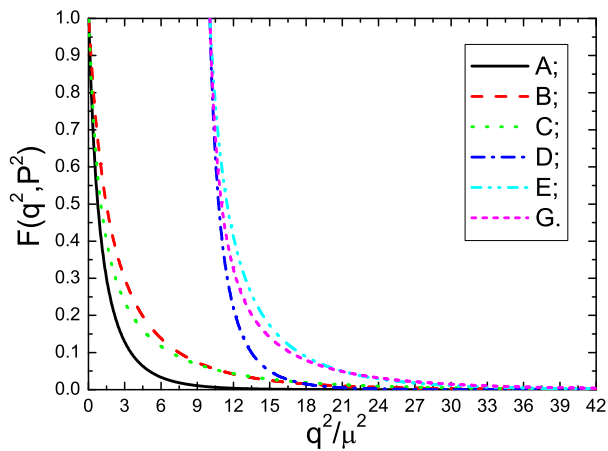}
  \caption{The Bethe-Salpeter amplitudes of the baryon-antibaryon bound states where $P^2=-M_{X}^2$, $A$, $B$, $C$, $D$, $E$ and $G$ denote the $p\bar{p}$,
  $\Sigma\bar{\Sigma}$, $\Xi\bar{\Xi}$,
$p\bar{N}(1440)$, $\Sigma\bar{\Sigma}(1660)$ and
$\Xi\bar{\Xi}^\prime$, respectively. The line-shapes   of the $D$,
$E$ and $G$ are plotted with the replacement  $q^2/\mu^2 \to
(q^2-10\mu^2)/\mu^2$ to avoid overlapping  with the $A$, $B$ and
$C$.}
\end{figure}

\begin{figure}
 \centering
 \includegraphics[totalheight=7cm,width=9cm]{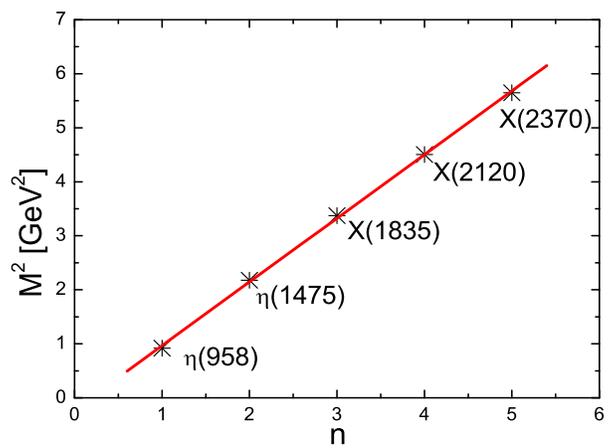}
  \caption{The Regge trajectory for the $\eta'$ mesons.  }
\end{figure}

The numerical results for the Bethe-Salpeter amplitudes are shown in
Fig.1 and the values of the bound states are presented in Table 1.
From the Table, we can see that the new resonances $X(1835)$ and
$X(2370)$ can be tentatively identified as the $p\bar{p}$ and
$p\bar{N}(1440)$ (or $N(1440)\bar{p}$) bound states, respectively,
while the mass of the $\Sigma\bar{\Sigma}$ bound state disfavors
identifying it as the $X(2120)$ or $X(2370)$, because the energy
gaps  $M_{\Sigma\bar{\Sigma}}-M_{X(2120)}=195.4\,\rm{MeV}$ and
$M_{\Sigma\bar{\Sigma}}-M_{X(2370)}=-58.5\,\rm{MeV}$.

In Ref.\cite{gb-Hao}, Hao, Qiao and Zhang  study the $0^{-+}$
three-gluon glueball with the  QCD sum rules, and  observe that its
mass lies in the region of $(1.9-2.7)\,\rm{GeV}$, while the quenched
lattice QCD calculations  indicate that the pseudoscalar glueballs
have masses about $2590 (40) (130)\,\rm{MeV}$ \cite{Latt-1-GB} or
$2560(35)(120)\,\rm{MeV}$ \cite{Latt-2-GB}. We should bear in mind
that it is the quenched lattice QCD not the full lattice QCD, as the
fermion determinant is neglected. In fact, the predictions vary with
the theoretical approaches, for example, the mass of the low-lying
pseudoscalar glueball $G$ is about $2.22\,\rm{GeV}$ from the
quenched QCD Hamiltonian in the Coulomb gauge  \cite{Swanson-GB},
$2.62\,\rm{GeV}$ from the constituent gluons model with the Cornell
potential within the helicity formalism \cite{Mathieu-GB},
$2.28\,\rm{GeV}$ from the string Hamiltonian derived from the vacuum
correlator method \cite{Simonov-GB}, $2.19\,\rm{GeV}$ from the
refined Gribov-Zwanziger version of the Landau gauge
\cite{Dudal-GB}, etc. In Ref.\cite{HYCheng-GB}, H. Y. Cheng  deduces
the mass of the pseudoscalar glueball $G$ from an $\eta$-$\eta'$-$G$
mixing formalism based on the anomalous Ward identity for transition
matrix elements, and  find a solution  $M_G=(1.4\pm 0.1)\,\rm{GeV}$
with the inputs from the KLOE experiment.  If the $\eta(1405)$ is
confirmed as a  pseudoscalar glueball one day, it is not necessary
that the low-lying pseudoscalar glueballs should have masses about
$2.5\,\rm{GeV}$. The new resonance $X(2120)$ may be a pseudoscalar
glueball, and there may be mixings among the baryonium states and
the glueballs, as the baryon-antibaryon bound states can annihilate
into three gluons.

In calculations, we have used a universal coupling constant $g^2$
($g=g^{\prime}$). In fact, no experiential data warrants that the
coupling constants $g_{NN\pi}$, $g_{NN(1440)\pi}$ and
$g_{N(1440)N(1440)\pi}$ have equal values. The experimental value
$\frac{g_{NN(1440)\pi}}{g_{NN\pi}}\approx 0.39-0.55$
\cite{Coupling1440-exp} and the theoretical value from a special
non-relativistic  quark model
$\frac{g_{NN(1440)\pi}}{g_{NN\pi}}\approx 0.33$
\cite{Coupling1440-the} deviate from $1$ obviously, we expect that
the values $|g_{N(1440)N(1440)\pi}|\leq |g_{NN\pi}|$. In
phenomenological applications, we often introduce the monopole (or
dipole) form-factors \cite{monopole-form} and the  exponential
form-factors \cite{exponential-form} to parameterize the off-shell
effects, and there are some form-factors associate with the coupling
constants. In this article, we use the modified Gaussian
distribution $g^2(k)=A
\left(\frac{k^2}{\mu^2}\right)^2\exp\left(-\frac{k^2}{\mu^2}\right)$,
which is assumed to take into account the form-factors effectively,
and introduce a parameter $\tau$ (with the value $0<\tau\leq 1$) to
parameterize the difference between the $g_{NN\pi}$ and the
$g_{N(1440)N(1440)\pi}$, i.e. $g^2(k)\rightarrow
g(k)g^{\prime}(k)=\tau A \left(\frac{k^2}{\mu^2}\right)^2\exp\left(-
\frac{k^2}{\mu^2}\right)$. We solve the Bethe-Salpeter equations
with variations of the parameters $\tau$ and $\mu$, where the flavor
$SU(3)$ breaking effects for the $\Sigma\bar{\Sigma}(1660)$ and
 $\Xi\bar{\Xi}'$  bound states are taken into account by the simple
replacements $\mu\rightarrow \mu \frac{M^2_{\Sigma}}{M_p^2}$ and $
\mu \frac{M^2_{\Xi}}{M_p^2}$ respectively, the eigenvalues are
presented in Table 2. In calculations, we observe  that in some
regions, there indeed exist solutions in the channels
$p\bar{N}(1440)$, $\Sigma\bar{\Sigma}(1660)$ and
$\Xi\bar{\Xi}^\prime$, and some eigenvalues of the $p\bar{N}(1440)$
(or $\bar{p}N(1440)$) bound state with $\tau=0.44-1.00$ and
$\mu=(200-400)\,\rm{MeV}$ are consistent with the mass of the
 $X(2370)$.

 The  radiative decays of the $J/\psi$  are generally
believed to be glue-rich, which can explain the branching ratio of
the decay $J/\psi \to \gamma\eta^\prime$  is large (about $(5.28 \pm
0.15) \times 10^{-3}$), while the branching  ratio of the $J/\psi
\to \gamma\eta$  is small  (about $( 1.104\pm0.034) \times10^{-3}$)
\cite{PDG}. The observation of the $X(1835)$, $X(2120)$ and
$X(2370)$ in the $\eta'$ channel not in the $\eta$ channel maybe due
to the intermediate virtual gluons  are flavor-neutral and the
$\eta'$ meson is mainly an $SU(3)$ flavor singlet and has
considerable gluon constituent via the axial anomaly. It is natural
to assume the $X(1835)$  and $X(2370)$ have some gluon constituents,
which play an important role in the decays to $\pi^+\pi^-\eta'$.

The hadronic molecular state $A$ which consists  of a meson pair or
a baryon pair $B+C$ can decay through two typical  routines, the
first one is $A\to B+C\to B+E+F+\cdots$, and the second one is $A\to
B+C\to E+F+G+\cdots$, then the decay widths are determined by the
intermediate process $C\to E+F$ or the annihilation of the $B+C$, we
can estimate the widths of the molecular states via  the decay
mechanisms. For example, in Ref.\cite{Guo4660}, Guo, Hanhart and
Meissner take the $Y(4660)$ as a $\psi'f_0(980)$ molecular state
considering   the nominal threshold of the $\psi'-f_0(980)$ system
is about $4666\pm10\,\rm{MeV}$ \cite{PDG}. The $Y(4660)$  decays
dominantly via the decay of the scalar meson $f_0(980)$, i.e.
$Y(4660)\to\psi'f_0(980)\to \psi'\pi\pi$, $ \psi'K{\bar K}$, and the
width of the $Y(4660)$ originates from the decay of the $f_0(980)$
mainly. On the other hand, if we take the $X(3872)$ as the
$D^*\bar{D}\pm\bar{D}^*D$ molecular state, the decay $X(3872)\to
\bar{D}^0D^0\pi^0$ can occur through the decays  $D^*\to D\pi^0$ and
$\bar{D}^*\to \bar{D}\pi^0$ \cite{Voloshin3872}, the narrow widths
of the  $D^*$ and $\bar{D}^*$ mesons warrant that the  width of the
$X(3872)$ is not broad, furthermore, the decay is suppressed
kinematically in the phase-space. In the present case, the
thresholds $2M_p=1876 \,\rm{MeV}>1835\,\rm{MeV}$ and
$M_p+M_{N(1440)}=2408\, \rm{MeV}>2376\,\rm{MeV}$,
 the decays $X(1835)\to p\bar{p}\to \eta'\pi^+\pi^-$ and $X(2370)\to p\bar{N}(1440),N(1440)\bar{p}\to \eta'\pi^+\pi^-$
can take place via  the Okubo-Zweig-Iizuka super-allowed fall apart
mechanism with re-arrangement in the color space.  The decays
$X(1835)\to p\bar{p} $ and $X(2370)\to p\bar{N}(1440),N(1440)\bar{p}
$ occur through the higher tails of the mass distributions, and the
widths may be large, although the decays are suppressed
kinematically at the lower tails of the mass distributions.  We can
search for the
 $ p\bar{N}(1440),N(1440)
 \bar{p}$ enhancements in the radiative decays $J/\psi \to \gamma p\bar{N}(1440), \gamma N(1440)
 \bar{p}$. The recent BESIII data indicates that the $X(1835)$ has
 the width $\Gamma=
(190 \pm 9^{+38}_{-36})\,\rm{MeV}$ \cite{BES-Shen,BES1012}, it is
too large for a pure molecular state. A larger mass glueball
constituent $G$ besides the $p\bar{p}$ component in the $X(1835)$ is
needed to take into account the experimental data, the decays  $G
\to ggg \to q\bar{q}q\bar{q}q\bar{q}$ can take place easily if
kinematically allowed, furthermore, such glue-rich processes prefer
the final-state $\eta^{\prime}\pi^+\pi^-$. The conventional
$q\bar{q}$ components in the $X(1835)$ and $X(2370)$ can also lead
to the decays $X(1835)\to p\bar{p} $ and $X(2370)\to
p\bar{N}(1440),N(1440)\bar{p} $ with the creation of additional  two
$q\bar{q}$ pairs from the QCD vacuum, however, the final-state
$\eta\pi^+\pi^-$ (rather than $\eta^{\prime}\pi^+\pi^-$) is
preferred as
 such processes are not glue-rich. It is difficult to calculate the
 decay widths of the $X(1835)$ and $X(2370)$ quantitatively in the
 framework of the Bethe-Salpeter equation.

If those bound states presented in Table 1 exist indeed, they can be
produced in the radiative $J/\psi$ decays, i.e.
$J/\psi\rightarrow\gamma gg$,
 $gg+q\bar{q}\rightarrow
p\bar{p}$, $\Sigma\bar{\Sigma}$, $\Xi\bar{\Xi}$,
 $p\bar{N}(1440)$,
$\Sigma\bar{\Sigma}(1660)$,  those bound states can decay to the
$\eta\pi\pi$, $\eta K\bar{K}$, $\eta'\pi\pi$, $\eta' K\bar{K}$,
$\eta'\eta\eta$, $\eta'\eta'\eta$, $\eta\eta\eta$ final states. We
can search for  those bound states in the $\eta\pi\pi$, $\eta
K\bar{K}$, $\eta'\pi\pi$, $\eta' K\bar{K}$, $\eta'\eta\eta$,
$\eta'\eta'\eta$, $\eta\eta\eta$ invariant mass distributions   in
the radiative decays  of the $J/\psi$ at the BESIII \cite{BESIII} or
the charmless $B$-decays at the KEK-B.

We do not exclude the canonical explanations, the $X(1835)$,
$X(2120)$ and $X(2370)$ may be the conventional mesons which
originate from the confining QCD forces and  consist of the
constituent quark-antiquark pairs with (or without) some gluon
components. In Ref.\cite{eta-Huang}, Huang and Zhu take the
$X(1835)$ as the second radial excited state of the $\eta'$, the
ground state nonet pseudoscalar  mesons are $\{\pi, K, \eta,
\eta'\}$, the first radial excited states are $\{\pi(1300), K(1460),
\eta(1295), \eta'(1475)\}$, and the second radial excited states are
$\{\pi(1800), K(1830), \eta(1760), X(1835)\}$. In
Ref.\cite{eta-Klempt},  Klempt and Zaitsev perform detailed analysis
of the properties of the $\eta(1295)$, $\eta(1405)$ and
$\eta(1475)$, and draw the conclusion that there maybe only one
$\eta$ state, the $\eta(1440)$, which has mass about
$(1200-1500)\,\rm{MeV}$, and identify the $X(1835)$ as the first
radial excited state of the $\eta'$. In Fig.2, we plot the $(n,M^2)$
for the $\eta^\prime$ mesons, where the $n$ denotes the radial
quantum numbers, the Regge trajectory favors   identifying  the
$\eta(1475)$, $X(1835)$, $X(2120)$ and $X(2370)$ as the radial
 excited $\eta^\prime$ mesons with $n=2$, $3$, $4$ and $5$,
respectively. The Regge trajectory alone cannot result in definite
identification. The decays $X(1835), X(2120), X(2370)\to
\eta'\pi^+\pi^-$ take place through the emission of a pair of
$S$-wave $\pi$ mesons, while the decays $X(1835), X(2120),
X(2370)\to \eta\pi\pi$ have not been observed experimentally yet.
Whether or not there exist those decay modes is of great importance,
further experiments are needed to prove or exclude the possibility.

\section{Conclusion}
In this article, we study the mass spectrum of the baryon-antibaryon
bound states $p\bar{p}$, $\Sigma\bar{\Sigma}$, $\Xi\bar{\Xi}$,
$\Lambda\bar{\Lambda}$, $p\bar{N}(1440)$,
$\Sigma\bar{\Sigma}(1660)$, $\Xi\bar{\Xi}^\prime$ and
$\Lambda\bar{\Lambda}(1600)$  in the framework of the Bethe-Salpeter
equation with a phenomenological potential. The numerical results
indicate that the $p\bar{p}$, $\Sigma\bar{\Sigma}$, $\Xi\bar{\Xi}$,
$p\bar{N}(1440)$, $\Sigma\bar{\Sigma}(1660)$, $\Xi\bar{\Xi}^\prime$
bound states maybe exist, and
 the new resonances $X(1835)$ and $X(2370)$ can be tentatively identified as the $p\bar{p}$
 and $p\bar{N}(1440)$ bound states respectively with some gluon constituents, while the new resonance
  $X(2120)$ may be a pseudoscalar glueball. The other  bound  states predicted in this work
   may be observed experimentally in
 the future in
the radiative decays  of the $J/\psi$ at the BESIII  or the
charmless $B$-decays at the KEK-B. On the other hand, the Regge
trajectory favors  identifying  the $\eta(1475)$, $X(1835)$,
$X(2120)$ and $X(2370)$ as the excited $\eta^\prime$ mesons with
$n=2$, $3$, $4$ and $5$, respectively.

\section*{Acknowledgments}
This  work is supported by National Natural Science Foundation,
Grant Number 11075053, and Program for New Century Excellent Talents
in University, Grant Number NCET-07-0282, and the Fundamental
Research Funds for the Central Universities.

\end{document}